\documentclass[aps,pra,amsmath,amssymb,tightenlines,epsfig,floatfix,twocolumn,superscriptaddress, longbibliography]{revtex4-1}
\usepackage{graphicx}
\usepackage{dcolumn}	% Align table columns on decimal point
\usepackage{bm}			% bold math
\usepackage{amsfonts}
\usepackage{xspace}
\usepackage{color}
\usepackage{epstopdf}
\usepackage{multirow}

\begin{document}

\newcommand{\psihat}{\ensuremath{\hat{\psi}}\xspace}
\newcommand{\psihatd}{\ensuremath{\hat{\psi}^{\dagger}}\xspace}
\newcommand{\ahat}{\ensuremath{\hat{a}}\xspace}
\newcommand{\Ham}{\ensuremath{\mathcal{H}}\xspace}
\newcommand{\ahatd}{\ensuremath{\hat{a}^{\dagger}}\xspace}
\newcommand{\bhat}{\ensuremath{\hat{b}}\xspace}
\newcommand{\bhatd}{\ensuremath{\hat{b}^{\dagger}}\xspace}
\newcommand{\boldr}{\ensuremath{\mathbf{r}}\xspace}
\newcommand{\dr}{\ensuremath{\,d^3\mathbf{r}}\xspace}
\newcommand{\tr}{\ensuremath{\,\mathrm{Tr}}\xspace}
\newcommand{\dk}{\ensuremath{\,d^3\mathbf{k}}\xspace}
\newcommand{\etal}{\emph{et al.\/}\xspace}
\newcommand{\ie}{i.e.\ }
\newcommand{\eq}[1]{Eq.\ (\ref{#1})\xspace}
\newcommand{\fig}[1]{Fig.\ \ref{#1}\xspace}
\newcommand{\abs}[1]{\left| #1 \right|}
\newcommand{\proj}[2]{\left| #1 \rangle\langle #2\right| \xspace}
\newcommand{\Qhat}{\ensuremath{\hat{Q}}\xspace}
\newcommand{\Qhatd}{\ensuremath{\hat{Q}^\dag}\xspace}
\newcommand{\phihatd}{\ensuremath{\hat{\phi}^{\dagger}}\xspace}
\newcommand{\phihat}{\ensuremath{\hat{\phi}}\xspace}
\newcommand{\boldk}{\ensuremath{\mathbf{k}}\xspace}
\newcommand{\boldp}{\ensuremath{\mathbf{p}}\xspace}
\newcommand{\boldsigma}{\ensuremath{\boldsymbol\sigma}\xspace}
\newcommand{\boldalpha}{\ensuremath{\boldsymbol\alpha}\xspace}
\newcommand{\grad}{\ensuremath{\boldsymbol\nabla}\xspace}
\newcommand{\parti}[2]{\frac{ \partial #1}{\partial #2} \xspace}
 \newcommand{\vs}[1]{\ensuremath{\boldsymbol{#1}}\xspace}
\renewcommand{\v}[1]{\ensuremath{\mathbf{#1}}\xspace}
\newcommand{\Psihat}{\ensuremath{\hat{\Psi}}\xspace}
\newcommand{\Psihatd}{\ensuremath{\hat{\Psi}^{\dagger}}\xspace}
\newcommand{\Vhatd}{\ensuremath{\hat{V}^{\dagger}}\xspace}
\newcommand{\Xhat}{\ensuremath{\hat{X}}\xspace}
\newcommand{\Xhatd}{\ensuremath{\hat{X}^{\dag}}\xspace}
\newcommand{\Yhat}{\ensuremath{\hat{Y}}\xspace}
\newcommand{\Jhat}{\ensuremath{\hat{J}}\xspace}
\newcommand{\Yhatd}{\ensuremath{\hat{Y}^{\dag}}\xspace}
\newcommand{\Uhat}{\ensuremath{\hat{U}^{\dag}}\xspace}
\newcommand{\jhat}{\ensuremath{\hat{J}}\xspace}
\newcommand{\lhat}{\ensuremath{\hat{L}}\xspace}
\newcommand{\Nhat}{\ensuremath{\hat{N}}\xspace}
\newcommand{\rhohat}{\ensuremath{\hat{\rho}}\xspace}
\newcommand{\ddt}{\ensuremath{\frac{d}{dt}}\xspace}
\newcommand{\nset}{\ensuremath{n_1, n_2,\dots, n_k}\xspace}
\newcommand{\Var}{\ensuremath{\mathrm{Var}}\xspace}
\newcommand{\Erf}{\ensuremath{\mathrm{Erf}}\xspace}

\newcommand{\notes}[1]{{\color{blue}#1}}
\newcommand{\sah}[1]{{\color{magenta}#1}}
	
\title{Using Interaction-Based Readouts to Approach the Ultimate Limit of Detection Noise Robustness for Quantum-Enhanced Metrology in Collective Spin Systems. }
\author{Simon A.~Haine}
\email{simon.a.haine@gmail.com}
\affiliation{Department of Physics and Astronomy, University of Sussex, Brighton BN1 9QH, United Kingdom}
\affiliation{Department of Quantum Science, Australian National University, Canberra, Australia}

\begin{abstract}
We consider the role of detection noise in quantum-enhanced metrology in collective spin systems, and derive a fundamental bound for the maximum obtainable sensitivity for a given level of added detection noise. We then present an interaction-based readout utilising the commonly used one-axis twisting scheme that approaches this bound for states generated via several commonly considered methods of generating quantum enhancement, such as one-axis twisting, two-axis counter-twisting, twist-and-turn squeezing, quantum non-demolition measurements, and adiabatically scanning through a quantum phase transition. We demonstrate that our method performs significantly better than other recently proposed interaction-based readouts. These results may help provide improved sensitivity for quantum sensing devices in the presence of unavoidable detection noise. 
\end{abstract}

\maketitle
There is a continued push for improved metrological potential in devices such as atomic clocks, atomic magnetometers, and inertial sensors based on atom interferometry \cite{Cronin:2009}. The physics of these systems is well described by collective spin-systems \cite{Pezze:2016_review}. Over the last decade there has been rapid progress in the demonstration of quantum enhanced metrology in these systems, that is, parameter estimation with sensitivity surpassing the shot-noise limit (SNL) \cite{Esteve:2008, Appel:2009, Leroux:2010, Schleier-Smith:2010, Schleier-Smith:2010b, Gross:2010, Riedel:2010, Lucke:2011, Hamley:2012, Berrada:2013, Ockeloen:2013, Strobel:2014, Muessel:2014, Muessel:2015, Krusse:2016, Hosten:2016, Zou:2018}. These schemes generally require a state preparation step, where inter-particle entanglement is created to enhance the metrological potential \cite{Hyllus:2010, Hyllus:2012, Toth:2012}, before the classical parameter of interest (which is usually proportional to a phase) is encoded onto the state.  There exists a plethora of state preparation techniques for creating highly quantum enhanced states, such as quantum state transfer from light to atoms \cite{Agarwal:1990,Kuzmich:1997,Moore:1999, Jing:2000, Fleischhauer:2002b, Haine:2005, Haine:2005b, Haine:2006b, Szigeti:2014b, Haine:2015}, quantum non-demolition measurement (QND), \cite{Kuzmich:1998, Kuzmich:2000, Appel:2009, Louchet-Chauvet:2010, Hammerer:2010, Hosten:2016}, spin changing collisions \cite{Duan:2000, Pu:2000, Lucke:2011, Hamley:2012, Nolan:2016}, one-axis twisting (OAT) \cite{Kitagawa:1993, Sorensen:2001, Esteve:2008, Gross:2010, Riedel:2010, Schleier-Smith:2010, Haine:2014}, two-axis counter-twisting (TACT) \cite{Kitagawa:1993, Ma:2009}, twist-and-turn squeezing (TNT) \cite{Law:2001, Muessel:2015}, and adiabatically scanning through a quantum phase transition (QPT) \cite{Lee:2006, Lee:2009, Zhang:2013, Xing:2016, Luo:2017, Feldmann:2018, Huang:2018}. However, the states generated via these schemes almost always require detection with very low noise (of the order of less than one particle) in order to see significant quantum enhancement \cite{Demkowicz-Dobrzanski:2012, Demkowicz-Dobrzanski:2014, Pezze:2016_review}. 

Recently, there has been considerable interest in the concept of interaction-based readouts (IBRs) \cite{Davis:2016, Hosten:2016b, Frowis:2016, Macri:2016, Linnemann:2016, Szigeti:2017, Nolan:2017b, Fang:2017, Anders:2018, Huang:2018, Feldmann:2018, Hayes:2018, Mirkhalaf:2018, Huang:2018b, Lewis-Swan:2018}, which are periods of unitary evolution applied to the system \emph{after} the phase encoding step, but \emph{before} the measurement takes place. These readouts usually involve inter-particle interactions, similar to the ones used for the state preparation.  Davis \etal showed that by using OAT to prepare a state with high quantum Fisher information (QFI), applying a phase shift, and then employing an IBR that reverses the OAT dynamics, quantum enhanced sensitivity could be achieved well beyond the Gaussian spin-squeezing regime. Furthermore, this quantum enhancement persisted even when the added detection noise was as large as the projection noise \cite{Davis:2016}. Similarly, Hosten \etal experimentally demonstrated that a period of nonlinear evolution after the state preparation and phase encoding could achieve sub SNL sensitivity in the presences of significant detection noise \cite{Hosten:2016b}. Macri \etal demonstrated that by performing an IBR that perfectly reverses the state preparation and then projects into the initial state, the sensitivity saturates the quantum Cram{\'e}r-Rao bound (QCRB) \cite{Macri:2016}. Nolan \etal \cite{Nolan:2017b} further generalised this result to show that there exist \emph{many} IBRs that satisfy the conditions for saturating the QCRB, and that the choice of IBR has implications for the level of sensitivity in the presence of detection noise (or ``robustness''). In particular, it was found that the optimum IBR was not necessarily the one that perfectly reversed the state preparation.  Furthermore, it was demonstrated that sensitivity approaching the Heisenberg limit \cite{Holland:1993, Giovannetti:2006} could be achieved in the presence of detection noise approaching the number of particles. IBRs have also been explored by applying time-reversal of the state-preparation dynamics in systems where the quantum-enhanced state is generated via SCC \cite{Gabbrielli:2015, Linnemann:2016, Szigeti:2017}, TACT \cite{Anders:2018}, TNT \cite{Mirkhalaf:2018}, and QPT \cite{Huang:2018,Feldmann:2018}.

In this work, we derive a limit for sensitivity in the presence of detection noise, which is significantly better than the levels achievable via previous schemes. We then present an IBR based on OAT that approaches this limit for states generated via OAT, TNT, TACT, QPT, and QND.

\section{Ultimate sensitivity limit in the presence of detection noise} 
\begin{figure}
%\centering 
\includegraphics[width =0.9\columnwidth]{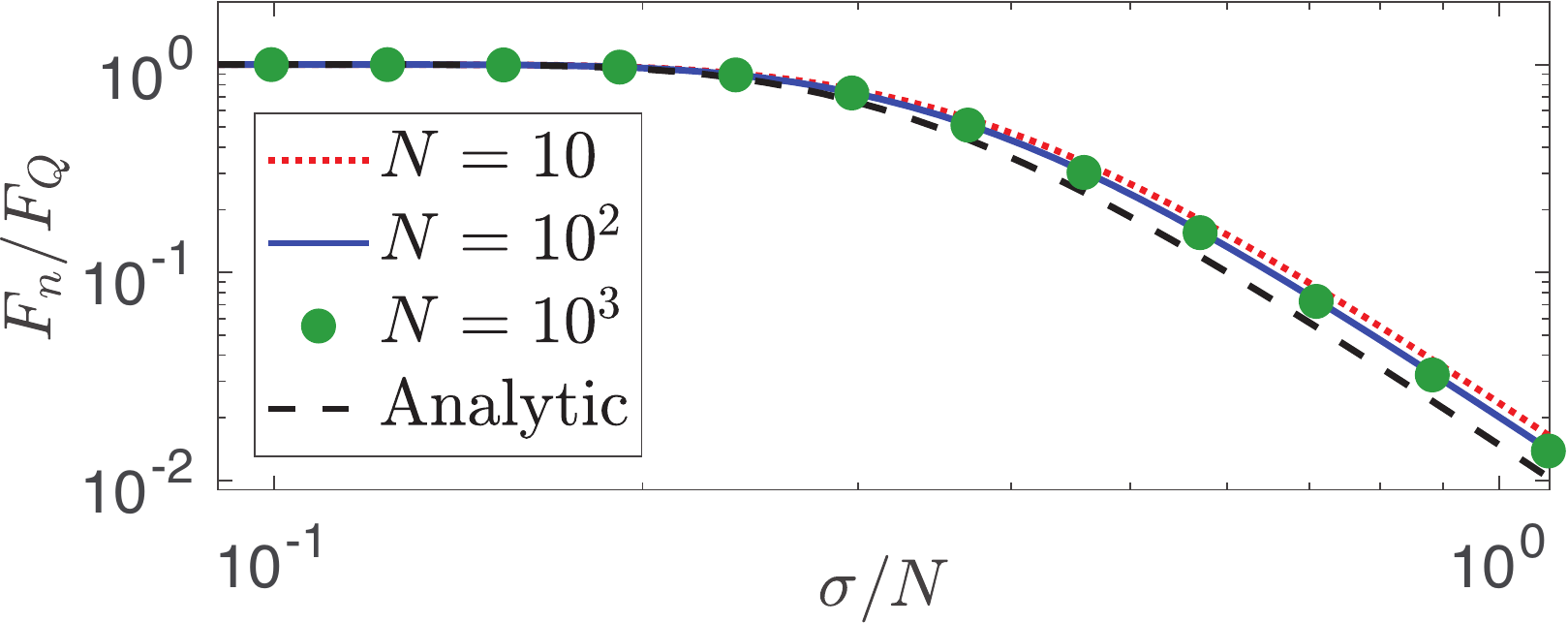}
\caption{The exact numeric value of $F_n$ vs.~$\sigma/N$ for $N = 10$, $10^2$ and $10^3$, compared to the approximate expression \eq{robust_limit}. The shape of $F_n(\sigma/N)$ is almost identical for $N=10^2$ and $10^3$.}
\label{fig:NQCRB}
\end{figure}

The sensitivity with which we can estimate the classical parameter $\phi$ is quantified via the Cram{\'e}r-Rao bound: $\Delta \phi^2 = 1/F_C$, where $F_C$ is the classical Fisher information (CFI), defined by $F_C = \sum_m \dot{P}_m^2/P_m$, where $P_m$ is the probability of obtaining measurement result $m$, and $\dot{P}_m \equiv \partial_\phi P_m$. Assuming a collection of $N$ particles distributed amongst two modes, the natural description for our system is provided via the \emph{pseudo-spin} SU(2) algebra: $[\Jhat_x, \Jhat_y] = i\Jhat_z$ \cite{Yurke:1986}. The eigenstates of these operators form a natural basis of easily accessible measurements, as they can be obtained via single-particle operations such as linear rotations and particle counting \cite{Pezze:2016_review}. For simplicity, throughout this paper we assume that measurements are made by projecting into the $\Jhat_z$ basis, \ie,  $\{\proj{m}{m}\}$, where $\Jhat_z|m\rangle = m |m\rangle$. The particular direction is of little consequence, however, as projections along other directions can be obtained via linear rotations. Following the convention introduced in \cite{Pezze:2013} and subsequently used in \cite{Gabbrielli:2015, Frowis:2016, Nolan:2017b, Pezze:2016_review, Feldmann:2018, Anders:2018, Mirkhalaf:2018, Huang:2018b}, we model the behaviour of an imperfect detector as sampling from the probability distribution 
\begin{equation}
\tilde{P}_m(\sigma) = \sum_{m^\prime} \Gamma_{m,m^\prime}(\sigma) P_{m^\prime} \, , \label{noise_def1}
\end{equation}
where 
\begin{equation}
\Gamma_{m, m^\prime}(\sigma) =   e^{-(m-m^\prime)^2/(2\sigma^2)}/\sum_m e^{-(m-m^\prime)^2/(2\sigma^2)}
\end{equation}
introduces detection noise of magnitude $\sigma$. This is equivalent to the positive operator valued measurement (POVM) $\{\hat{M}_m\} = \{\sum_{m^\prime}\Gamma_{m, m^\prime}|m^\prime\rangle\langle m^\prime | \}$. To demonstrate how the noise affects the CFI, we consider the case where $P_m$ contains only two non-zero elements, $P_a$ and $P_b$, with $P_b =1-P_a$, and $\dot{P}_a = -\dot{P}_b =  \sqrt{F_0(P_a-P_a^2)}$, such that $F_C = F_0$.  By approximating $m$ as a continuous variable and extending the domain to $\pm \infty$  \footnote{See the supplemental material for further details of the derivation of Eq.~(\ref{robust_limit}).}, we obtain 
\begin{equation}
\tilde{P}(m) = (P_a e^{-(a-m)^2/2\sigma^2} + P_b e^{-(b-m)^2/2\sigma^2})/\sqrt{2\pi}\sigma \, .
\end{equation}
Defining 
\begin{equation}
\tilde{P}_{a} = \int_{-\infty}^{\tfrac{1}{2}(a+b)} \tilde{P}(m) dm \quad \mathrm{and}\quad \tilde{P}_{a} = \int_{\tfrac{1}{2}(a+b)}^\infty \tilde{P}(m) dm \label{P_ab}
\end{equation}
(assuming $a < b$), and maximising with respect to $P_a$ ($P_a\rightarrow P_b \rightarrow \frac{1}{2}$)  we obtain 
\begin{equation}
F_C(\sigma) = \dot{\tilde{P}}^2_a/\tilde{P}_a + \dot{\tilde{P}}^2_b/\tilde{P}_b \approx F_0 \left(\Erf\left[(a-b)/2\sqrt{2}\sigma\right]\right)^2 \, .
\end{equation}
Clearly, $F_C(\sigma)$ decays less rapidly when the separation between the non-zero components of $P_m$, $|a-b|$, is large compared to $\sigma$. This intuition leads us to postulate that distribution with maximum robustness, $P_\mathrm{opt}$ is 
\begin{subequations}
\begin{align}
P_{N/2} &= P_{-N/2} = \frac{1}{2}, \\
\dot{P}_{N/2} &= -\dot{P}_{-N/2} = \sqrt{F_0}/2, 
\end{align}
\end{subequations}
with all other elements equal to zero. While an analytic proof of this remains elusive, we confirm this via a numeric optimisation method \footnote{See the supplemental material for further details of the optimisation method}. In the absence of detection noise, the QCRB states that $F_C\leq F_Q$, where $F_Q$ is the QFI. We define the \emph{noisy QCRB} (NQCRB) as $F_C(\sigma) \leq F_{n}(\sigma)$, where $F_{n}(\sigma)$ is the CFI calculated from the $\{\tilde{P}_m(\sigma), \dot{\tilde{P}}_m(\sigma)\}$ obtained from performing the discrete sum in \eq{noise_def1} numerically with $\{P_m, \dot{P}_m\} =  \{P_\mathrm{opt}, \dot{P}_\mathrm{opt}\}$, and setting $F_0 = F_Q$. This is the maximum sensitivity that can be achieved by making spin measurements on a state with QFI equal to $F_Q$ in the presence of detection noise $\sigma$. We can get an approximate analytic expression for $F_{n}(\sigma)$ by again approximating $m$ as a continuous variable, but limiting the range to $-N/2 < m < N/2$, such that
\begin{equation}
F_{n}(\sigma) \approx F_Q\left(1- 2 \frac{\Erf [\alpha/2]}{\Erf[\alpha]}\right)^2 \, , \label{robust_limit}
\end{equation}
with $\alpha = N/\sqrt{2} \sigma$. Fig.(\ref{fig:NQCRB}) shows excellent agreement between this expression and the exact value of $F_{n}(\sigma)$, calculated numerically. \eq{robust_limit} provides a slight under-estimate of the CFI, as information is lost when condensing $P_m$ into a binary distribution via \eq{P_ab}. For the remainder of this paper, we use the exact numeric value of $F_{n}(\sigma)$ rather than \eq{robust_limit}. 

\begin{figure*}
%\centering 
\includegraphics[width =\textwidth]{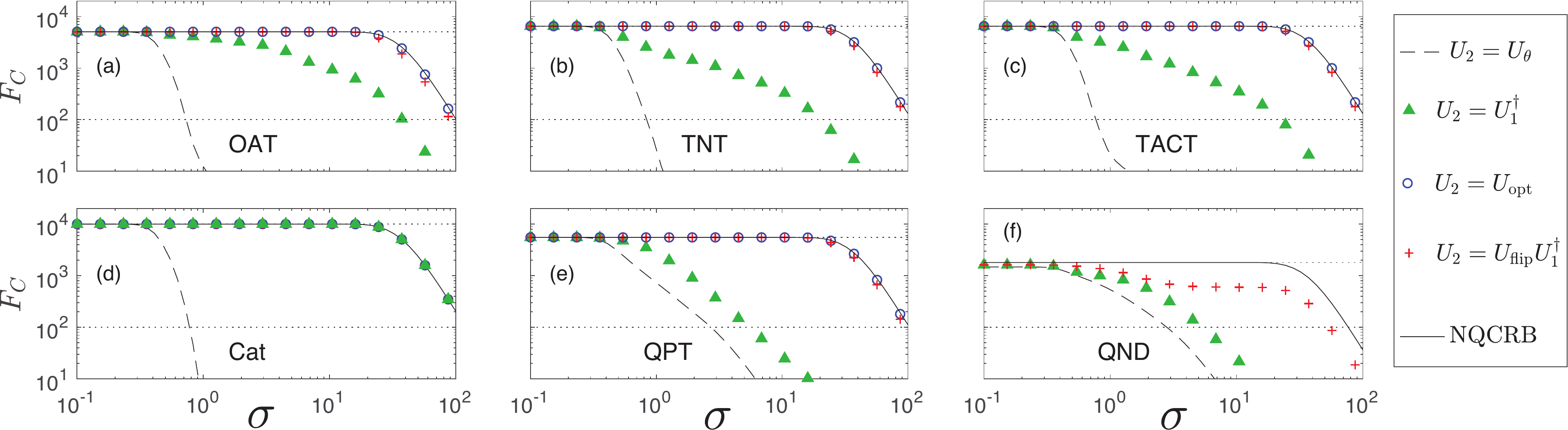}
\caption{$F_C(\sigma)$ for (a): OAT with $r=0.2$, (b): TNT, (c): TACT, (d): OAT with $r=\tfrac{\pi}{2}$ (which corresponds to a spin-cat state), (e): QPT, and (f): QND. $U_\theta = e^{i\tfrac{\pi}{2} \Jhat_y}$ for OAT, Cat, and TNT, and $U_\theta = 1$ for TACT, QPT, and QND. The upper and lower dotted black lines indicate the QCRB ($F_C = F_Q$) and SNL ($F_C = N$), respectively. $N=100$ for all cases, and we have optimised over $\phi$. The optimum $\phi$ is close to $\phi_0$ for $U_2 = U_\mathrm{opt}$, and close to $0$ for $U_2 = U_\mathrm{flip}U_1^\dag$. }
\label{fig:fig1}
\end{figure*}

\section{Interaction-based readout to saturate the NQCRB} 
\begin{figure}
%\centering 
\includegraphics[width =1\columnwidth]{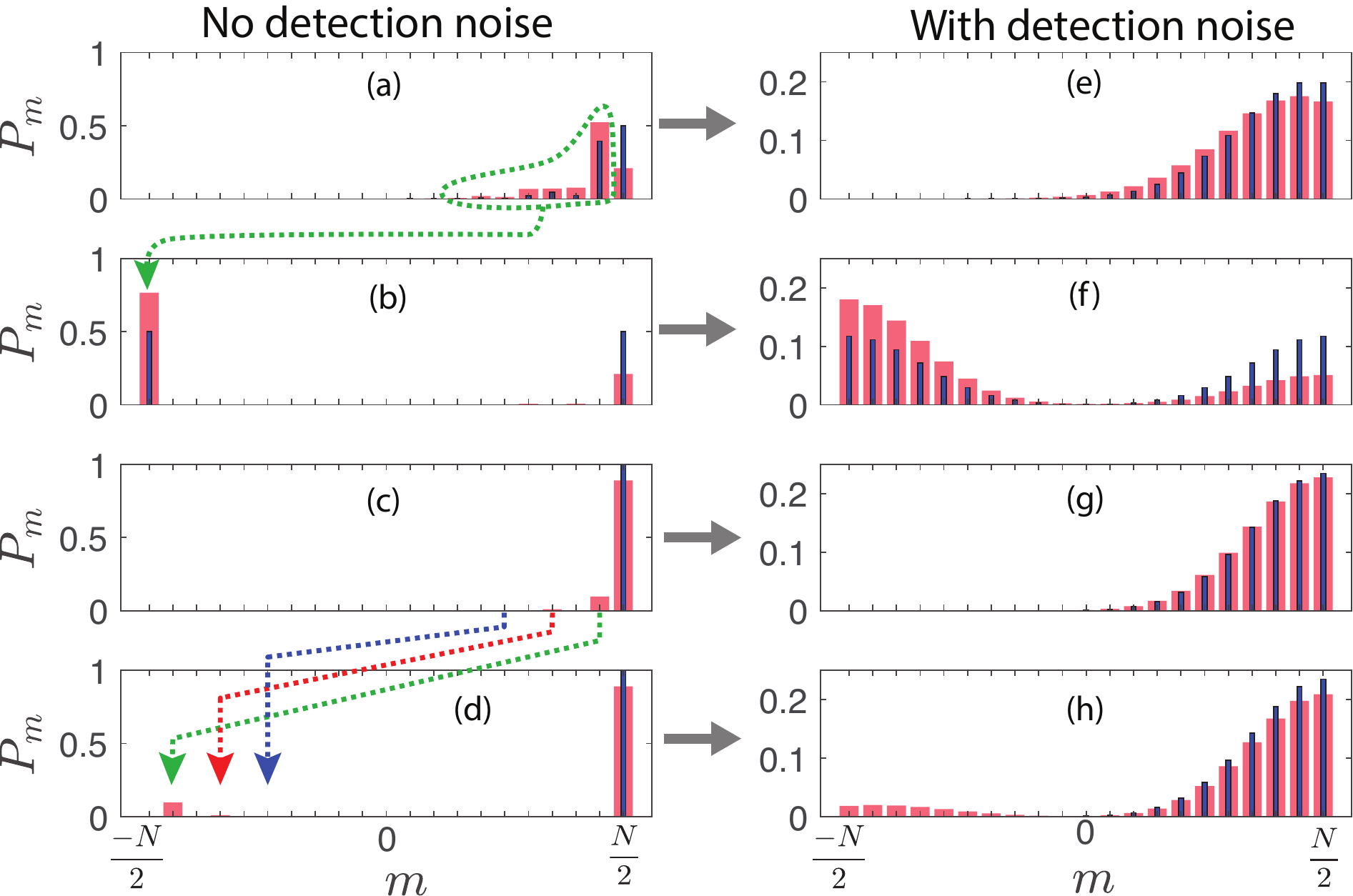}
\caption{ $P_m(\phi)$ (blue thin bars) and $P_m(\phi+\delta \phi)$ (pink thick bars) with (right column) and without (left column) detection noise $\sigma$. (a) \& (e): $U_2 = U_1^\dag$, $\phi = \phi_0$. (b)\&(f): $U_2 = U_\mathrm{opt}$, $\phi = \phi_0$. (c)\&(g): $U_2 = U_1^\dag$, $\phi = 0$. (d)\&(h) $U_2 = U_\mathrm{flip}U_1^\dag$, $\phi = 0$. The Hellinger distance $d_H$ is  (a-d): $0.238$, (e): $0.067$,  (f): $0.201$, (g): $0.012$, (h): $0.232$. Parameters: $N=20$, $\sigma = 3$, $r=0.2$, $\delta_\phi = \tfrac{1}{N}$, $\phi_0 = 0.118$. The behaviour of $U_\mathrm{p}$ and $U_\mathrm{flip}$ is indicated by the arrows between (a)\&(b), and (c)\&(d), respectively.}
\label{fig:probs}
\end{figure}

The NQCRB sets the maximum achievable CFI in the presence of detection noise $\sigma$. What remains is to find an IBR that allows us to achieve this limit. Starting with an arbitrary initial pure state $|\psi_1\rangle$, we note that this state can always be written as $|\psi_1\rangle = U_1|\psi_0\rangle$, where  $|\psi_0\rangle = |\tfrac{N}{2}\rangle$ is the maximal $\Jhat_z$ eigenstate, which is completely separable in the particle basis. In most quantum enhanced metrology schemes, the unitary operator $U_1$ implements the \emph{state preparation} step, which may be employed to increase the QFI of an initially separable state. Specific examples of this process including OAT, TACT, TNT, and QPT will be considered later. The phase shift $\phi$ is then encoded on to the state via $|\psi_\phi\rangle = e^{i \Jhat_n \phi}|\psi_1\rangle$, where $\Jhat_n = \mathbf{J}\cdot \mathbf{n}$, and $\mathbf{n}$ is a unit vector chosen to maximise the QFI of $|\psi_\phi\rangle$. This vector can be obtained from the collective covariance matrix \cite{Hyllus:2010}. An IBR is some unitary $U_2$ such that measurements are made on the state $U_2|\psi_\phi\rangle$. Our goal is to find $U_2$ such that the probability distribution $P_m = |\langle m|U_2 |\psi_\phi\rangle|^2$ saturates the NQCRB. It was shown in \cite{Macri:2016} that for $\phi \ll 1$, selecting $U_2 = U_1^\dag$ saturates the QCRB. At some value $\phi = \phi_0$, 
\begin{equation}
U_1^\dag e^{i \Jhat_n \phi_0}U_1|\psi_0\rangle = \frac{1}{\sqrt{2}}(|\psi_0\rangle + |\psi^\prime\rangle) \equiv |\psi_b \rangle \, ,
\end{equation}
where 
\begin{equation}
|\psi^\prime\rangle = (\hat{1} - \proj{\psi_0}{\psi_0})|\psi_{b}\rangle/\sqrt{1-\abs{\langle \psi_{b}|\psi_0\rangle}^2} \, .
\end{equation}
We can artificially construct an IBR that is maximally robust to noise simply by constructing a unitary operator $U_\mathrm{p}$ that maps this state to one with distribution $P_\mathrm{opt}$: 
\begin{equation}
U_\mathrm{p} = |\tfrac{N}{2}\rangle\langle \tfrac{N}{2} | + |\tfrac{-N}{2} \rangle\langle \psi^\prime |  + \sum_{m=-N/2+1}^{N/2-1} \proj{m}{m^\prime} ,
\end{equation}
where $\{|m^\prime\rangle\}$ completes the orthogonal basis containing $|\frac{N}{2}\rangle$ and $|\psi^\prime\rangle$. Thus, the optimum IBR is 
\begin{equation}
U_2 = U_\mathrm{p} U_1^\dag \equiv U_\mathrm{opt}. 
\end{equation}
Fig.~(\ref{fig:fig1}) shows the CFI calculated from $P_m = |\langle m|U_\mathrm{opt} |\psi_\phi\rangle|^2$ after convolving with detection noise, for quantum enhanced states generated from OAT, TACT, TNT, and QPT. Details of these states are provided in table (\ref{tab1}) \footnote{For further details on these quantum states, see the supplemental material. By `quantum-enhanced states', we mean `states with $F_Q > N$'. }. 
\begin{table}[h]
\begin{center}
\begin{tabular}{| l || l | l|}
Scheme: & $U_1$ & $ r$ \\
\hline
OAT & $e^{i r \Jhat_z^2}e^{i\tfrac{\pi}{2} \Jhat_y}$ & $0.2 $ \\
TACT & $e^{i r(\Jhat_x^2 - \Jhat_y^2)}$ & $0.032 $ \\
TNT & $ e^{i r (\Jhat_z^2 - \tfrac{N}{2}\Jhat_x)}e^{i\tfrac{\pi}{2} \Jhat_y}$ & $0.0715 $ \\
Cat & $e^{i r \Jhat_z^2} e^{i\tfrac{\pi}{2} \Jhat_y}$ & $\tfrac{\pi}{2} $ \\
QPT & $ \mathcal{T} \exp \left( \tfrac{-i}{\hbar}\int_0^{t_0} \hat{H}(t^\prime) dt^\prime \right) e^{i\tfrac{\pi}{2} \Jhat_y}$ &
\end{tabular}
\end{center}
\caption{Details of the quantum state $|\psi_1\rangle = U_1 |\tfrac{N}{2}\rangle$ used in Fig. (\ref{fig:fig1}). For TACT and TNT, $r$ was chosen to maximise $F_Q$ for $N=100$, while for OAT, a moderate value of $r$ was chosen such that the state was no longer in the spin-squeezed regime \cite{Davis:2016}, but not sufficient to reach the maximum QFI spin-cat state, which occurs at $r=\tfrac{\pi}{2}$.}
\label{tab1}
\end{table}%
In all cases, we find that this IBR saturates the NQCRB. To understand the mechanism for this, we consider the effect of detection noise on the probability distributions. Fig.\ (\ref{fig:probs}) shows $P_m(\phi)$ and $P_m(\phi+\delta \phi)$, with (right column) and without (left column) noise, for the case of OAT. When $U_2 = U_1^\dag$ ((a) and (e)), the change in probability is centred around $m= \frac{N}{2}$ and nearby elements. When detection noise is added, $P_m(\phi)$ and $P_m(\phi + \delta \phi)$ become less distinct as the adjacent elements are mixed. However, by applying $U_2 = U_\mathrm{opt}$ ((b) and (f)), all of the probability in elements $m\neq \tfrac{N}{2}$ is transferred to $m = -\tfrac{N}{2}$ such that $P_m = P_\mathrm{opt}$. We stress that the application of $U_\mathrm{opt}$ does not effect the CFI in the absence of noise - the Hellinger distance 
\begin{equation}
d_H^2 = 1-\sum_m \sqrt{P_m(\phi)P_m(\phi+\delta \phi)}
\end{equation}
is identical in (a) and (b) ($d_H \approx 0.24$). However, $U_\mathrm{opt}$ \emph{does} effect how distinguishable the states remain after the addition of detection noise: $d_H \approx 0.067$, and $0.201$ for (e) and (f) respectively. 

\section{Approaching the NQCRB with OAT-based IBRs } 
While our optimum IBR gives us insight into what maximises robustness, it is of no use to us unless we can find a physical mechanism with which it can be implemented. However, we can construct an IBR which has similar properties to the ideal case with the OAT  mechanism. The OAT unitary can be used to create the well known spin-cat state \cite{Agarwal:1997, Nolan:2017}: 
\begin{equation}
e^{i\tfrac{\pi}{2} \Jhat_y^2}|m\rangle = \tfrac{1}{\sqrt{2}}e^{i\tfrac{\pi}{4}}(|m\rangle +i(-1)^m|-m\rangle) \equiv |\beta(m)\rangle \, ,
\end{equation}
for even $N$ \footnote{For odd $N$ we require an additional rotation: an equal superposition cat is generated by $e^{i\frac{\pi}{2}\hat{J}_y}e^{i\frac{\pi}{2} \hat{J}_y^2}|m\protect\rangle$.}. This state has the unusual property that $| \langle \beta(m) |e^{i\tfrac{\pi}{2}\Jhat_z} |\beta(m)\rangle |^2 = \cos^2 \frac{m\pi}{2}$. That is, even-$m$ states are unaffected by a $\pi$ rotation, while odd-$m$ states become orthogonal. As such, a $\tfrac{\pi}{2}$ phase shift followed by secondary application of $e^{i\tfrac{\pi}{2} \Jhat_y^2}$ will return $|\beta(m)\rangle$ to $|m\rangle$ if $m$ is even, or transfer it to an orthogonal state if $m$ is odd. Specifically
\begin{align}
e^{i\tfrac{\pi}{2}\Jhat_y^2}e^{i\tfrac{\pi}{2} \Jhat_z}e^{i\tfrac{\pi}{2}\Jhat_y^2} = -\sum_m i^{m(m-1)} |-1^m m\rangle\langle m| \equiv U_\mathrm{flip} \, .
\end{align}
The action of $U_\mathrm{flip}$ is to exchange the odd elements of $P_m$ with $P_{-m}$, while leaving the even elements unaffected, as illustrated in fig.(\ref{fig:probs}) (d) and (h) \footnote{For odd $N$, an IBR that performs the same function is given by $U_\mathrm{flip} = e^{i\tfrac{\pi}{2}\Jhat_y(\Jhat_y+1)}e^{i\theta \Jhat_z}e^{i\tfrac{\pi}{2}\Jhat_y(\Jhat_y+1)}$, with $\theta =\frac{\pi}{2}(1+1/N)$}. For sufficiently small $\phi$, most of the CFI for the state $U_1^\dag |\psi_\phi\rangle$ is usually contained in the elements $m=\tfrac{N}{2}$ and $m=\tfrac{N}{2}-1$ ((c) and (g)). Applying $U_\mathrm{flip}$ to this state transfers probability from $m=\tfrac{N}{2}-1$ to $m=-(\tfrac{N}{2}-1)$, forming a distribution \emph{almost} as robust as $P_\mathrm{opt}$. 

Fig.\ (\ref{fig:fig1}) shows the performance of this scheme compared to $U_\mathrm{opt}$ for quantum enhanced states generated via OAT, TACT, and TNT (see table (\ref{tab1})). In these three cases we see that $U_2 = U_\mathrm{flip}U_1^\dag$ is very close to the optimum case ($U_2 = U_\mathrm{opt}$ and the NQCRB), and achieves sensitivity very close to the QCRB for detection noise $\sigma$ significantly exceeding $\sqrt{N}$. For comparison, we have also included the previously considered case of an \emph{echo}, where $U_2 = U_1^\dag$ , which performs significantly better than the case of no IBR ($U_2 = U_\theta$, where only a linear rotation is used to maximise the CFI), but not nearly as well as $U_2 = U_\mathrm{flip}U_1^\dag$. We have also included the special case of OAT with $r = \tfrac{\pi}{2}$, which corresponds to the maximum QFI spin-cat state. In this case, both $U_2= U_\mathrm{flip}U_1^\dag$ and $U_2= U_1^\dag$ saturate the NQCRB, while the case of no IBR loses all quantum enhancement for $\sigma \lessapprox 1$. The reason why there is no need for the extra application of $U_\mathrm{flip}$ is because the state $U_1^\dag|\psi_\phi\rangle$ already yields a probability distribution identical to $P_\mathrm{opt}$, and is unchanged by application of $U_\mathrm{flip}$. The outstanding performance of the echo IBR for this state was first reported in \cite{Nolan:2017b} and subsequently in \cite{Fang:2017, Huang:2018b}, but it was not known that this is the maximum achievable sensitivity \footnote{We note that \cite{Fang:2017} reports higher robustness than this. However, the state is identical, and the discrepancy is due to a different convention for the detection noise}. 

We also considered QPT, where the increased QFI is generated by slowly varying the parameters in a time-dependent Hamiltonian, such that the ground state is adiabatically transformed to one with high QFI. We implemented this with a Hamiltonian of the form 
\begin{equation}
\hat{H} = \hbar \chi (\Jhat_x \cos^2 \tfrac{\pi}{2} \tfrac{t}{t_0} + \Jhat_z^2 \sin^2 \tfrac{\pi}{2} \tfrac{t}{t_0}),
\end{equation} 
such that 
\begin{equation}
U_1 = \mathcal{T} \left[ \exp \left( \tfrac{-i}{\hbar}\int_0^{t_0} \hat{H}(t^\prime) dt^\prime \right)\right]e^{i \tfrac{\pi}{2}\Jhat_y}, 
\end{equation}
where $\mathcal{T}$ represents the time-ordering operator. In the limit $\chi t_0\rightarrow \infty$, $U_1|\tfrac{N}{2}\rangle = |0\rangle$, the twin-Fock state. We chose a moderate value $\chi t_0 = 20$, such that the final state contains non-zero elements on either side of  $m=0$. Unlike the previous examples, when making measurements on the state $U_1^\dag |\psi_\phi\rangle$ for small $\phi$, most of the CFI is contained in the elements $m = \tfrac{N}{2}$ and $m = \tfrac{N}{2}-2$, such that $U_\mathrm{flip}$ has little effect. This is easily rectified, however, by using a modified IBR with $U_\mathrm{flip}^\prime = e^{i\tfrac{\pi}{2}\Jhat_y^2}e^{i\tfrac{\pi}{4} \Jhat_z}e^{i\tfrac{\pi}{2}\Jhat_y^2}$, which for $N \gg 1$, $U_\mathrm{flip}|m\rangle \approx |-m\rangle$ if $m/2$ is odd. We see in Fig.\ (\ref{fig:fig1}e) that this IBR is very close to the NQCRB. 

The benefit of our IBR is not limited to pure states. We consider a quantum enhanced mixed state 
\begin{equation}
\rho = \sum_m e^{-\frac{m^2}{\Delta^2}}|m\rangle\langle m| /(\sum_m e^{-\tfrac{m^2}{\Delta^2}}). 
\end{equation}
We chose $\Delta =1$, which corresponds to a state with significant quantum enhancement, yet is far from pure, with the purity $\gamma = \mathrm{Tr}[\rho^2] \approx 0.4$. Such a state may arise from quantum enhancement via a strong QND interaction with a detuned optical field, as described in \cite{Haine:2015b}, with an imperfect measurement leading to uncertainty in $m$. Unlike the previous states considered, this state is mixed, so there is no unitary operator that maps this distribution to $P_\mathrm{opt}$. However, at $\phi=0$, the final distribution is similar to the QPT case, which inspires us to use the same IBR, namely $U_2 = U_\mathrm{flip} U_1^\dag$, with $U_1$ generated via the adiabatic evolution considered in the QPT example. We see in Fig.\ (\ref{fig:fig1}f) that while this case isn't as robust as previous examples, the general trend is the same, that is $U_2 = U_\mathrm{flip} U_1^\dag$ is more robust than $U_2 = U_1^\dag$, which in turn outperforms $U_2 =U_\theta$. As the state is mixed, we cannot systematically construct $U_\mathrm{opt}$.  For completeness, we have also investigated applying our IBR to states with no quantum enhancement, such as coherent spin-states \cite{Radcliffe:1971}, and find qualitatively similar results \footnote{The plot of $F_C(\sigma)$ for the coherent spin state is provided in the supplementary material}. 

\section{Discussion} The results of this paper may form an integral part of future quantum-enhanced sensing technologies, as high-QFI states are particularly susceptible to detection noise. While OAT-based quantum enhancement schemes are not yet capable of manufacturing spin-cat states (and therefore $U_\mathrm{flip}$), progress in this area is rapid, particularly in schemes based on optically induced non-linearities \cite{Schleier-Smith:2010, Hosten:2016}, and Rydberg atoms \cite{Busche:2017}. Furthermore, we have provided insight and a systematic approach for constructing a robust IBR. Armed with this insight, schemes that approximate our optimum scheme may be found through other dynamical mechanisms that are perhaps easier to implement in a particular system. For example, it has been shown that QPT can be used to engineer spin-cat states \cite{Lee:2009}, so could potentially be used to construct a near-optimum IBR. One might question the wisdom of using an IBR that requires the ability to create a maximum QFI cat state in cases where the QFI of the input state is less than this. However, there may be situations when it is impractical to use a state preparation capable of creating a cat state, such as when the preparation time is limited \cite{Hayes:2018}. Similarly, a state with less quantum enhancement may be desirable in the presence of external phase noise. In these situations, the presence of unavoidably large detection noise will still necessitate the use of a high-performance IBR in order to achieve high sensitivity. Finally, the NQCRB provides a limit for the performance of all IBR's. Once the sensitivity approaches this limit, further gains can only be made through the reduction of detection noise, rather than via improvement of the IBR. 

\begin{acknowledgements}
The author acknowledges fruitful discussions with Samuel Nolan, Safoura Mirkhalaf, Luca Pezze, Augusto Smerzi, Manuel Gessner, and Jacob Dunningham. This work was supported by the European Union's Horizon 2020 research and innovation programme under the Marie Sklodowska-Curie grant agreement No.~704672.  
\end{acknowledgements}

\bibliography{../../../simon_bib.bib}

\newpage

\begin{widetext}
\section*{Supplemental Material}
In this supplemental material I provide further details about the derivation of the noisy quantum Cram{\'e}r-Rao bound (NQCRB), and provide further details about the quantum states used in this manuscript. 

\section{Derivation of Eq.~(5)}
Beginning with Eq.~(1),  
\begin{equation}
\tilde{P}_m = \sum_{m^\prime} \Gamma_{m,m^\prime}(\sigma) P_{m^\prime} \, , \label{noise_def_sup1}
\end{equation}
and Eq.~(2),
\begin{equation}
\Gamma_{m, m^\prime}(\sigma) =   \frac{e^{-(m-m^\prime)^2/(2\sigma^2)}}{\sum_m e^{-(m-m^\prime)^2/(2\sigma^2)}} \, , \label{noise_def_sup2}
\end{equation}
we can obtain an approximate expression for the case when $P_m$ contains only two non-zero elements, at $m=a$ and $m=b$. 
By treating the discrete probability distribution as continuous, we obtain
\begin{equation}
P(m) = P_a \delta(m-a) + P_b\delta(m-b) \, .
\end{equation}
Replacing the discrete sum in \eq{noise_def_sup1} with a continuous integral, we find
\begin{equation}
\tilde{P}(m) = \int_{-\infty}^\infty \Gamma(m-m^\prime) P(m^\prime) dm
\end{equation}
where
\begin{equation}
\Gamma(m-m^\prime)  = \frac{1}{\sigma \sqrt{2\pi}} e^{-(m-m^\prime)^2/(2\sigma^2)} \, .
\end{equation}
Defining 
\begin{subequations}
\begin{eqnarray}
\tilde{P}_{a} &=& \int_{-\infty}^{\tfrac{1}{2}(a+b)} \tilde{P}(m) dm \\ 
\tilde{P}_{b} &=& \int_{\tfrac{1}{2}(a+b)}^\infty \tilde{P}(m) dm
\end{eqnarray}
\end{subequations}
gives
\begin{subequations}
\begin{eqnarray}
\tilde{P}_{a} &=& \tfrac{1}{2}\left(1 + (2P_a-1 )\Erf \left[\frac{b-a}{2\sqrt{2}\sigma} \right]\right) \\
\tilde{P}_{b} &=& \tfrac{1}{2}\left(1 + (1 - 2P_a)\Erf \left[\frac{b-a}{2\sqrt{2}\sigma} \right]\right)
\end{eqnarray}
\end{subequations}
where we have used $P_b = 1-P_a$. Similarly, we find
\begin{subequations}
\begin{eqnarray}
\dot{\tilde{P}}_{a} &=& \dot{P}_a\Erf \left[\frac{b-a}{2\sqrt{2}\sigma} \right] \\
\dot{\tilde{P}}_{b} &=& - \dot{P}_a\Erf \left[\frac{b-a}{2\sqrt{2}\sigma} \right]
\end{eqnarray}
\end{subequations}
where we have used $\dot{P}_b = -\dot{P}_a$. Using these equations in $F_C(\sigma) = \dot{\tilde{P}}_a^2/\tilde{P}_a + \dot{\tilde{P}}_b^2/\tilde{P}_b$ gives
\begin{equation}
F_C(\sigma) = \frac{4 \dot{P}_a^2 \Erf \left[ \frac{b-a}{2\sqrt{2}\sigma}\right]^2}{(1-2P_a)^2 \Erf \left[ \frac{b-a}{2\sqrt{2}\sigma}\right]^2 -1}
\end{equation}
Setting $\dot{P}_a = \sqrt{F_0(P_a-P_a^2)}$, such that 
\begin{equation}
F_C(\sigma =0) = \frac{\dot{P}_a^2}{P_a} + \frac{\dot{P}_b^2}{P_b} = F_0 \, ,
\end{equation}
gives
\begin{equation}
F_C(\sigma) = F_0\frac{4P_a(1- P_a) \Erf \left[ \frac{b-a}{2\sqrt{2}\sigma}\right]^2}{1-(1-2P_a)^2 \Erf \left[ \frac{b-a}{2\sqrt{2}\sigma}\right]^2} \, .
\end{equation}
Maximising this function with respect to $P_a$ (Setting $\partial_{P_a} F_C(\sigma) = 0$ and solving for $P_a$) gives $P_a = \tfrac{1}{2}$, and therefore
\begin{equation}
F_C(\sigma) = F_0\Erf \left[ \frac{b-a}{2\sqrt{2}\sigma}\right]^2 \, .
\end{equation}

\section{Optimum probability distribution in the presence of detection noise}
In this section we demonstrate that of all probability distributions with $F_C = F_0$, $P_\mathrm{opt}$, the distribution with $P_{\tfrac{N}{2}} = P_{-\tfrac{N}{2}} = \tfrac{1}{2}$, $\dot{P}_{\tfrac{N}{2}} = -\dot{P}_{-\tfrac{N}{2}} = \sqrt{F_0}/2$, displays the maximum sensitivity in the presence of detection noise $\sigma$. We introduce the vectors
\begin{subequations}
\begin{eqnarray}
\v{v} &=& \left[\sqrt{P_{-\tfrac{N}{2}}}, \sqrt{P_{-\tfrac{N}{2}+1}}, \dots, \sqrt{P_{\tfrac{N}{2}-1}},\sqrt{P_{\tfrac{N}{2}}}\right]^T \\
\dot{\v{v}} &=& \partial_\phi \left[\sqrt{P_{-\tfrac{N}{2}}}, \sqrt{P_{-\tfrac{N}{2}+1}}, \dots, \sqrt{P_{\tfrac{N}{2}-1}},\sqrt{P_{\tfrac{N}{2}}}\right]^T \nonumber \\
&=& \frac{1}{2}\left[\frac{\dot{P}_{-\tfrac{N}{2}}}{\sqrt{P_{-\tfrac{N}{2}}}}, \frac{\dot{P}_{-\tfrac{N}{2}+1}}{\sqrt{P_{-\tfrac{N}{2}+1}}}, \dots, \frac{\dot{P}_{\tfrac{N}{2}-1}}{\sqrt{P_{\tfrac{N}{2}-1}}},\frac{\dot{P}_{\tfrac{N}{2}}}{\sqrt{P_{\tfrac{N}{2}}}}\right]^T \, ,
\end{eqnarray}
\end{subequations}
such that 
\begin{equation}
F_C = 4\dot{\v{v}}^T \dot{\v{v}} = \sum_m \frac{\dot{P}_m^2}{P_m} \, .
\end{equation}
Using this notation, its straightforward to transform our distribution such that $\v{v}^\prime = A\v{v}$, $\dot{\v{v}}^\prime = A\dot{\v{v}}$, where $A$ is a square orthogonal real matrix with the property $A^T A = A A^T = 1$. Importantly, such a transformation preserves the CFI: 
\begin{eqnarray}
F_C(\v{v}^\prime) &=& 4 (\dot{\v{v}}^\prime)^T \dot{\v{v}}^\prime \nonumber \\
&=&  4\left(A\dot{\v{v}}\right)^T\left(A \dot{\v{v}}\right)\nonumber \\
&=&  4\dot{\v{v}}^T A^T  A \dot{\v{v}} \nonumber \\
&=& 4\dot{\v{v}}^T \dot{\v{v}}  = F_C(\v{v}) \, .
\end{eqnarray}
To confirm that $P_\mathrm{opt}$ is in fact the distribution with maximum robustness, we begin with an arbitrary probability distribution $P_\mathrm{arb}$ that satisfies $F_C\left[P_\mathrm{arb}, \dot{P}_\mathrm{arb}\right] = F_0$, and then employ a numeric optimisation algorithm, which is implemented as follows:
\begin{enumerate}
\item Calculate $\{\v{v}, \dot{\v{v}}\}$ from $\{P_m, \dot{P}_m\}$. 
\item Rotate $\v{v}$ and $\dot{\v{v}}$ by a small angle of randomly generated magnitude about a randomly generated axis in $N+1$ dimensional space. This process is represented by an orthogonal real matrix $A$, and therefore conserves $F_C(\sigma=0)$. 
\item Calculate $\{P_m, \dot{P}_m\}$ from the new $\{\v{v}, \dot{\v{v}}\}$. 
\item Add detection noise to this new distribution via \eq{noise_def_sup1} and \eq{noise_def_sup2}, and calculate $F_C(\sigma)$. If the $F_C(\sigma)$ has increased, we accept this new distribution, and repeat. Otherwise, we keep the original distribution, and repeat. 
\end{enumerate}  
Fig.~\ref{suppfig1} (a-c) shows the CFI after addition of detection noise for $10^5$ iterations of this algorithm, for three different initial distributions, all with $\dot{P}_m$ chosen such that $F_C(0) = 1$. However, each distribution has a different CFI in the presence of noise. The CFI (with detection noise) rapidly converges to the CFI of $P_\mathrm{opt}$. The evolution of the Hellinger distance between these distributions and $P_\mathrm{opt}$ approaches zero (d-f). We repeated this process for several different values of $\sigma$ and initial distributions, and in all cases found convergence to $P_\mathrm{opt}$. 

\begin{figure}[h]
	\begin{center}
		\includegraphics[width=\textwidth]{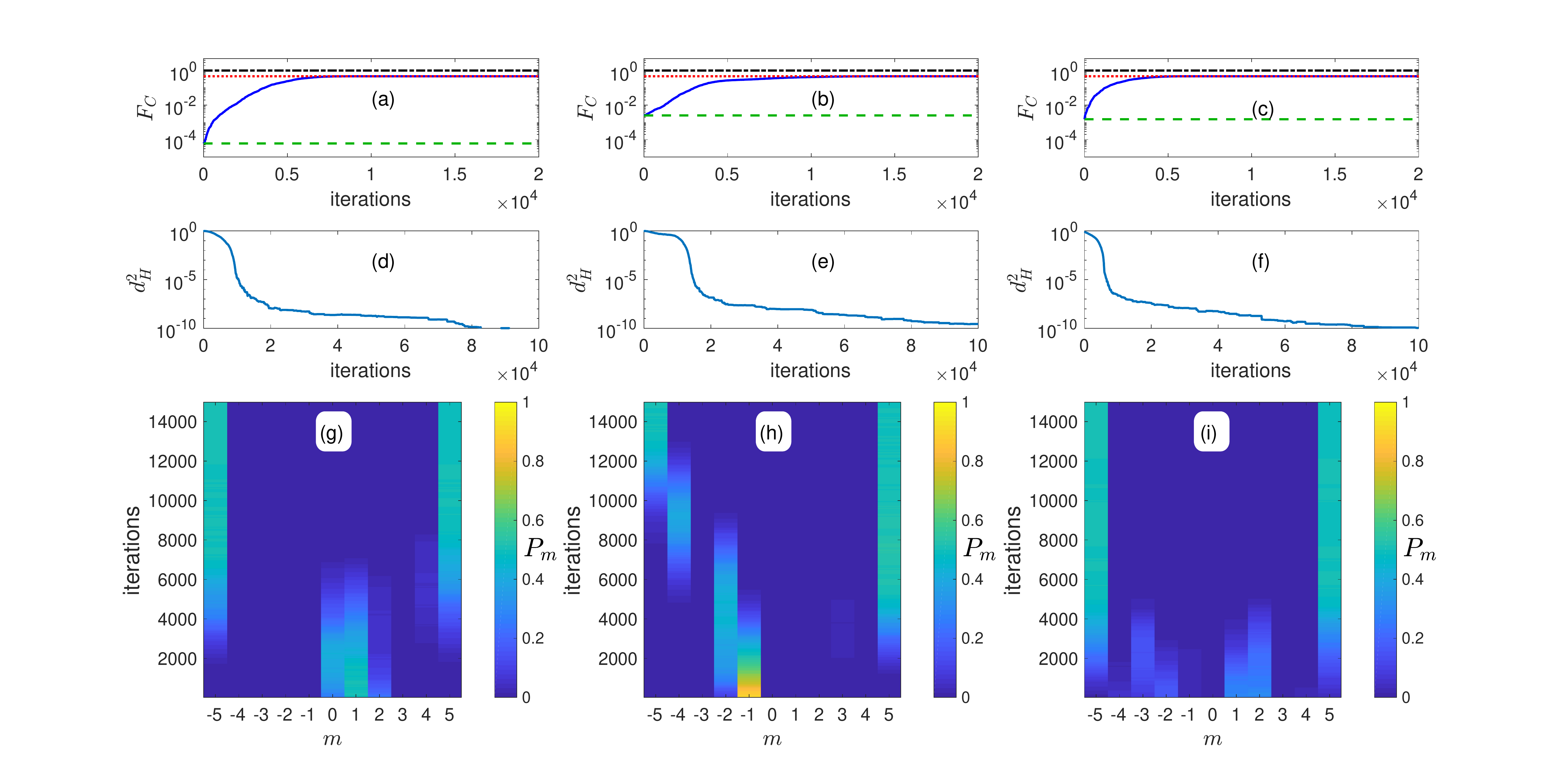}
		\caption{(a-c): The CFI in the presence of detection noise $\sigma$, $F_C(\sigma)$ as a function of the number of iterations of the optimisation algorithm (blue line), compared to the initial value (green dashed line). We have also show $F_C(\sigma)$ for $P_\mathrm{opt}$, which is the NQCRB (red dotted line). The CFI without noise ($F_C(0)$) remains constant for all iterations, and is identical to the $F_C(0)$ for $P_\mathrm{opt}$ (black dot dashed line). (d-f): The Hellinger distance between $P$ and $P_\mathrm{opt}$ vs. the number of iterations. (g-i): The evolution of the probability distributions corresponding to the above frames. Parameters: $N=10$, $\sigma = 4$.}
		\label{suppfig1}
	\end{center}
\end{figure}

To ensure that our optimisation algorithm is not getting `stuck' in a local maximum, we generate entirely random distributions satisfying the constraint that $F_C\left[P, \dot{P}\right] = F_0$, by employing a randomly generated transformation matrix to $P_\mathrm{opt}$. We see in Fig.\ (\ref{suppfig2}) that while $F_C(0)$ remains constant,  $F_C(\sigma)$ does not exceed the optimum value, calculated from $P_\mathrm{opt}$.  Again, we employed different initial distributions and values of $\sigma$. 

\begin{figure}[h]
	\begin{center}
		\includegraphics[width=\textwidth]{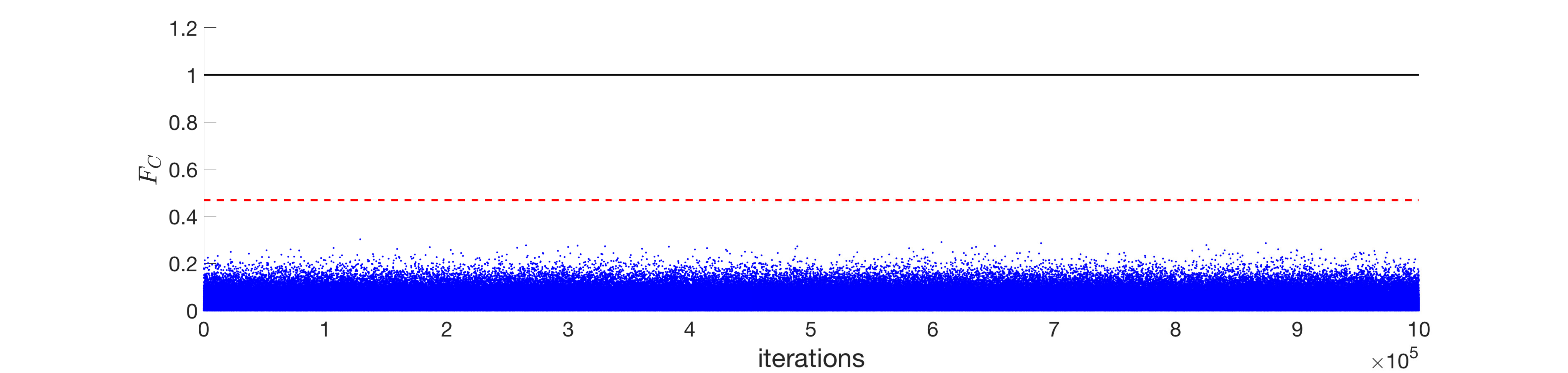}
		\caption{Blue dots: $F_C(\sigma)$ for randomly generated distributions with $F_C(0) =1$. $F_C(\sigma)$ does not exceed the NQCRB, indicated by the red dashed line. Parameters: $N=10$, $\sigma=4$.}
		\label{suppfig2}
	\end{center}
\end{figure}

\section{Derivation of Equation 6}
As before, we approximate $P_\mathrm{opt}$ as a continuous distribution such that
\begin{eqnarray}
P(m) &=& \tfrac{1}{2}\left(\delta(m + \tfrac{N}{2}) + \delta(m - \tfrac{N}{2})\right) \\
\dot{P}(m) &=&   \frac{\sqrt{F_0}}{2}\left(\delta(m + \tfrac{N}{2}) - \delta(m - \tfrac{N}{2})\right).
\end{eqnarray}
To derive equation (5) we made the approximation that the domain of integration extended to infinity, which is reasonable as long as $|a|,\, |b| \ll N/2$. However, in order to get a more accurate approximation, we now restrict our domain to $\{-N/2, N/2\}$. Approximating $\Gamma_{m,m^\prime}$ as a continuous function, and enforcing the correct normalisation conditions gives
\begin{eqnarray}
\tilde{P}(m) &=& \int_{-N/2}^{N/2} P(m^\prime) \Gamma(m-m^\prime) dm^\prime \nonumber \\
&=&\frac{1}{\sigma \Erf \left[\frac{N}{\sqrt{2}\sigma} \right]}\left(\exp\left( \frac{\left(m-\tfrac{N}{2}\right)^2}{2\sigma^2}\right) + \exp\left( \frac{\left(m+\tfrac{N}{2}\right)^2}{2\sigma^2}\right)\right) \\
\dot{\tilde{P}}(m) &=& \int_{-N/2}^{N/2} P(m^\prime) \Gamma(m-m^\prime) dm^\prime \nonumber \\
&=&\frac{\sqrt{F_0}}{\sigma \Erf \left[\frac{N}{\sqrt{2}\sigma} \right]}\left(\exp\left( \frac{\left(m-\tfrac{N}{2}\right)^2}{2\sigma^2}\right) - \exp\left( \frac{\left(m+\tfrac{N}{2}\right)^2}{2\sigma^2}\right)\right) \, .
\end{eqnarray}
Defining $\tilde{P}_a$ and $\tilde{P}_b$ as before, we find
\begin{subequations}
\begin{eqnarray}
\tilde{P}_a &=& \int_{-N/2}^0 \tilde{P}(m) dm = \frac{1}{2}\\
\tilde{P}_b &=& \int_{0}^{N/2} \tilde{P}(m) dm = \frac{1}{2}\\
\end{eqnarray}
\end{subequations}
and
\begin{subequations}
\begin{eqnarray}
\dot{\tilde{P}}_a &=& \int_{-N/2}^0 \dot{\tilde{P}}(m) dm 
= \sqrt{F_0}\left(\frac{1}{2} - \frac{\Erf\left[N/2\sqrt{2}\sigma \right]}{\Erf\left[N/\sqrt{2}\sigma \right]}\right) \\
\dot{\tilde{P}}_b &=& \int_{0}^{N/2} \dot{\tilde{P}}(m) dm 
=\sqrt{F_0}\left(-\frac{1}{2} + \frac{\Erf\left[N/2\sqrt{2}\sigma \right]}{\Erf\left[N/\sqrt{2}\sigma \right]}\right) \, .
\end{eqnarray}
\end{subequations}
Using these equations in $F_C(\sigma) = \dot{\tilde{P}}^2_a/P_a + \dot{\tilde{P}}^2_b/P_b$ gives
\begin{equation}
F_{C}(\sigma) = F_0\left(1- 2 \frac{\Erf [N/2\sqrt{2}\sigma]}{\Erf[N/\sqrt{2}\sigma]}\right)^2 \, .
\end{equation}
If we choose our IBRO such that the measurement saturates the QCRB in the absence of noise, we replace $F_0$ with $F_Q$, and arrive at equation (6) of the main text. 

\section{Further details of the quantum states used in figure 2}
In this section we give further details about the states used in figure (2) of the main text. We have used the Husimi $Q$-function as a visualisation tool, defined by 
\begin{equation}
Q(\theta, \phi) = \frac{N+1}{4\pi} \langle \theta, \phi | \rho |\theta, \phi\rangle
\end{equation}
with $\rho = \proj{\psi_1}{\psi_1}$, and
\begin{equation}
|\theta, \phi\rangle = \exp(i\phi \Jhat_z)\exp(i\theta \Jhat_y)|\frac{N}{2}\rangle \, .
\end{equation}
Additionally, $N=100$ was used throughout. 
\subsection{OAT}
The OAT state is generated via $|\psi_1\rangle = U_1 |\tfrac{N}{2}\rangle$, where 
\begin{equation}
U_1 = \exp\left(i r \hat{J}_z^2\right)\exp\left(i \tfrac{\pi}{2} \hat{J}_y\right). 
\end{equation}
For figure (2), we chose $r=0.2$. Fig.~\ref{oatfig} shows the QFI, probability distribution, and Husimi Q-Function.

\begin{figure}[h!!!]
	\begin{center}
		\includegraphics[width=\textwidth]{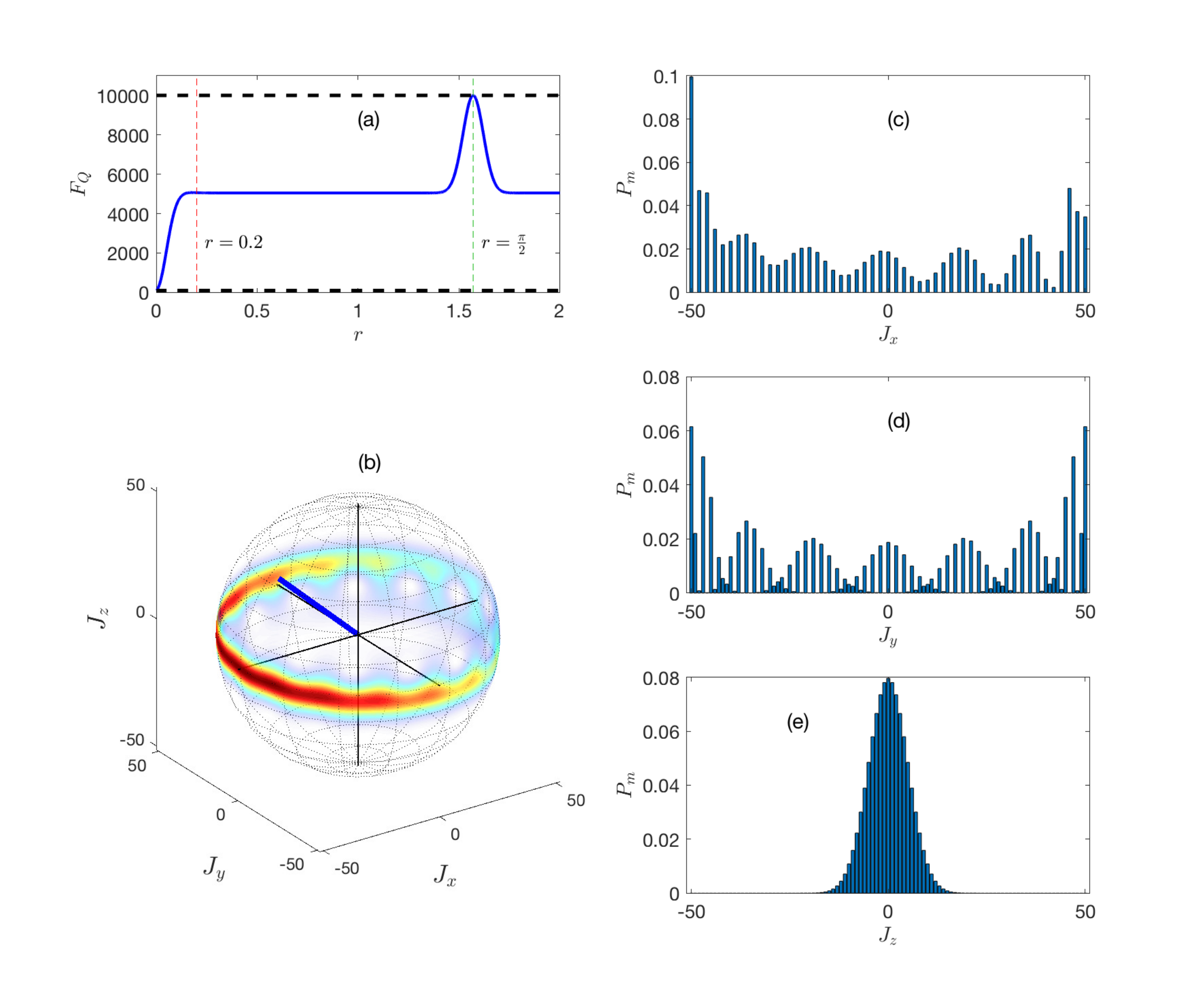}
		\caption{Properties of the OAT state: $|\psi_1\rangle = \exp(i r \hat{J}_z^2)\exp(i \tfrac{\pi}{2} \hat{J}_y)|\tfrac{N}{2}\rangle$. (a): The QFI is calculated via $F_Q = 4 \left(\langle \psi_1| \Jhat_n^2|\psi_1\rangle - |\langle \psi_1| \Jhat_n|\psi_1\rangle |^2\right)$, where $\Jhat_n$ is the pseudo-spin operator along which the QFI is maximum. The value of $r$ used in figure (2) ($r=0.2$) is indicated by the vertical red line. The lower and upper dashed black lines indicate the shot-noise limit ($F_Q = N$) and Heisenberg limit ($F_Q = N^2$), respectively.  (b) $Q(\theta, \phi)$. The direction of $\hat{J}_n$ is indicated by the thick blue line. (c-e): The probability distribution $P_m = |\langle m_j|\psi_1\rangle|^2$, where $m_j$ is the $m$th eigenstate of $\Jhat_j$, for $j = \{x,y,z\}$. }
		\label{oatfig}
	\end{center}
\end{figure}

\subsection{TNT}
The TNT state is generated via $|\psi_1\rangle = U_1 |\tfrac{N}{2}\rangle$, where 
\begin{equation}
U_1 = \exp\left(i r \left(\hat{J}_z^2-\frac{N}{2}\Jhat_x\right)\right)\exp(i \tfrac{\pi}{2} \hat{J}_y). 
\end{equation}
For figure (2), we chose $r=0.0715$, which is the value at which the QFI is maximum. The Husimi $Q$-function, probability distribution, and QFI for this state are shown in Fig.~\ref{TNTfig}. 

\begin{figure}[h!!!]
	\begin{center}
		\includegraphics[width=\textwidth]{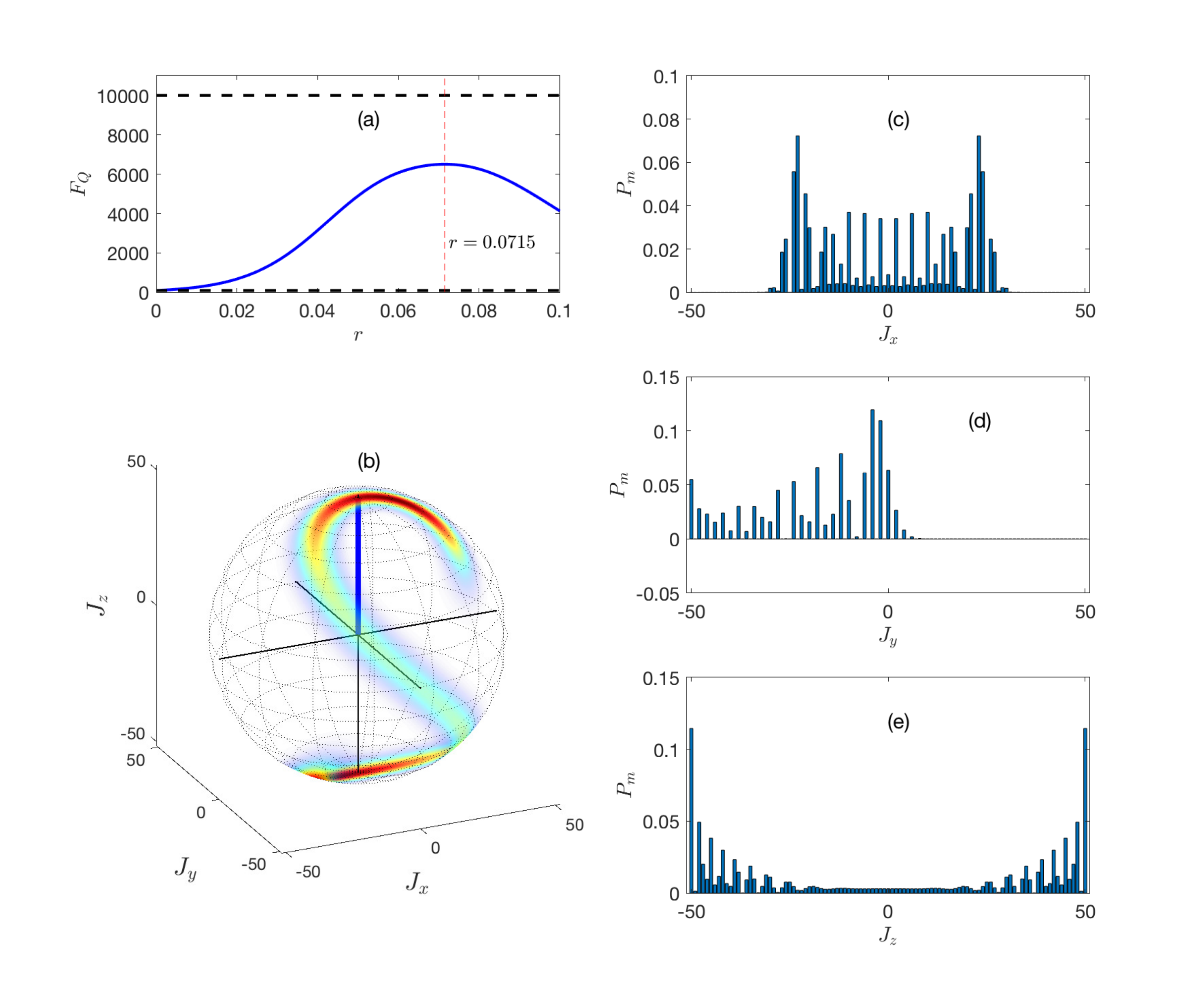}
		\caption{Properties of the TNT state: $|\psi_1\rangle = \exp\left(i r \left(\hat{J}_z^2-\frac{N}{2}\Jhat_x\right)\right)\exp(i \tfrac{\pi}{2} \hat{J}_y) |\tfrac{N}{2}\rangle$. (a): The QFI is calculated via $F_Q = 4 \left(\langle \psi_1| \Jhat_n^2|\psi_1\rangle - |\langle \psi_1| \Jhat_n|\psi_1\rangle |^2\right)$, where $\Jhat_n$ is the pseudo-spin operator along which the QFI is maximum. The value of $r$ used in figure (2) ($r=0.0715$) is indicated by the vertical red line. The lower and upper dashed black lines indicate the shot-noise limit ($F_Q = N$) and Heisenberg limit ($F_Q = N^2$), respectively.  (b) $Q(\theta, \phi)$. The direction of $\hat{J}_n$ is indicated by the thick blue line. (c-e): The probability distribution $P_m = |\langle m_j|\psi_1\rangle|^2$, where $m_j$ is the $m$th eigenstate of $\Jhat_j$, for $j = \{x,y,z\}$. }
		\label{TNTfig}
	\end{center}
\end{figure}

\subsection{TACT}
The TACT state is generated via $|\psi_1\rangle = U_1 |\tfrac{N}{2}\rangle$, where 
\begin{equation}
U_1 = \exp\left(i r \left(\hat{J}_x^2-\Jhat_y^2\right)\right). 
\end{equation}
For figure (2), we chose $r=0.032$, which is the value at which the QFI is maximum. The Husimi $Q$-function, probability distribution, and QFI for this state are shown in Fig.~\ref{tatfig}. 

\begin{figure}[h!!!]
	\begin{center}
		\includegraphics[width=\textwidth]{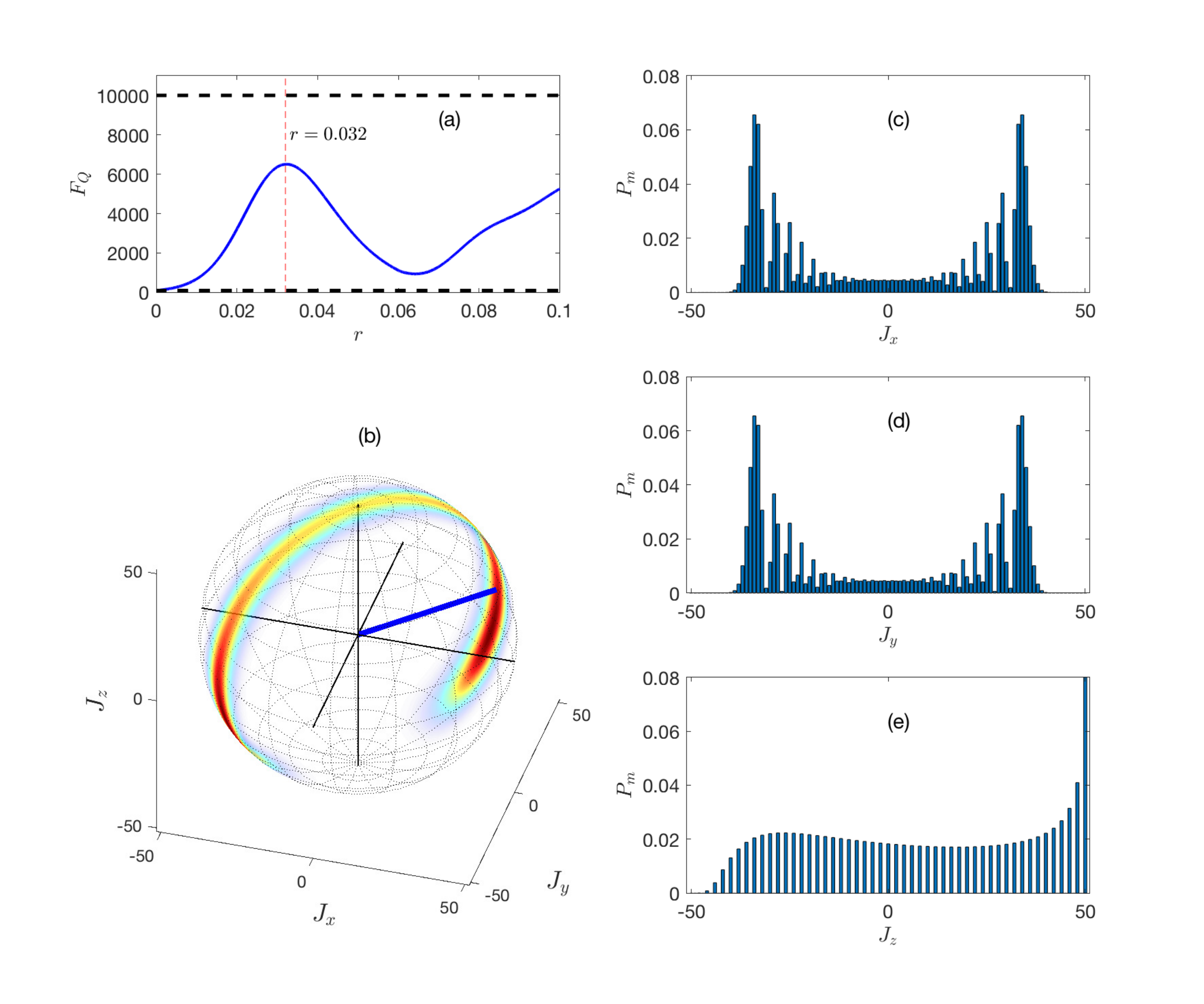}
		\caption{Properties of the TACT state: $|\psi_1\rangle = \exp\left(i r \left(\hat{J}_x^2-\Jhat_y^2\right)\right) |\tfrac{N}{2}\rangle$. (a): The QFI is calculated via $F_Q = 4 \left(\langle \psi_1| \Jhat_n^2|\psi_1\rangle - |\langle \psi_1| \Jhat_n|\psi_1\rangle |^2\right)$, where $\Jhat_n$ is the pseudo-spin operator along which the QFI is maximum. The value of $r$ used in figure (2) ($r=0.032$) is indicated by the vertical red line. The lower and upper dashed black lines indicate the shot-noise limit ($F_Q = N$) and Heisenberg limit ($F_Q = N^2$), respectively.  (b) $Q(\theta, \phi)$. The direction of $\hat{J}_n$ is indicated by the thick blue line. (c-e): The probability distribution $P_m = |\langle m_j|\psi_1\rangle|^2$, where $m_j$ is the $m$th eigenstate of $\Jhat_j$, for $j = \{x,y,z\}$. }
		\label{tatfig}
	\end{center}
\end{figure}

\subsection{Cat}
The cat state is generated via $|\psi_1\rangle = U_1 |\tfrac{N}{2}\rangle$, where 
\begin{equation}
\exp\left(i r \hat{J}_z^2\right)\exp\left(i \tfrac{\pi}{2} \hat{J}_y\right). 
\end{equation}
with $r=\pi/2$, which is the value at which the QFI is maximum. The Husimi $Q$-function, probability distribution, and QFI for this state are shown in Fig.~\ref{catfig}. 

\begin{figure}[h!!!]
	\begin{center}
		\includegraphics[width=\textwidth]{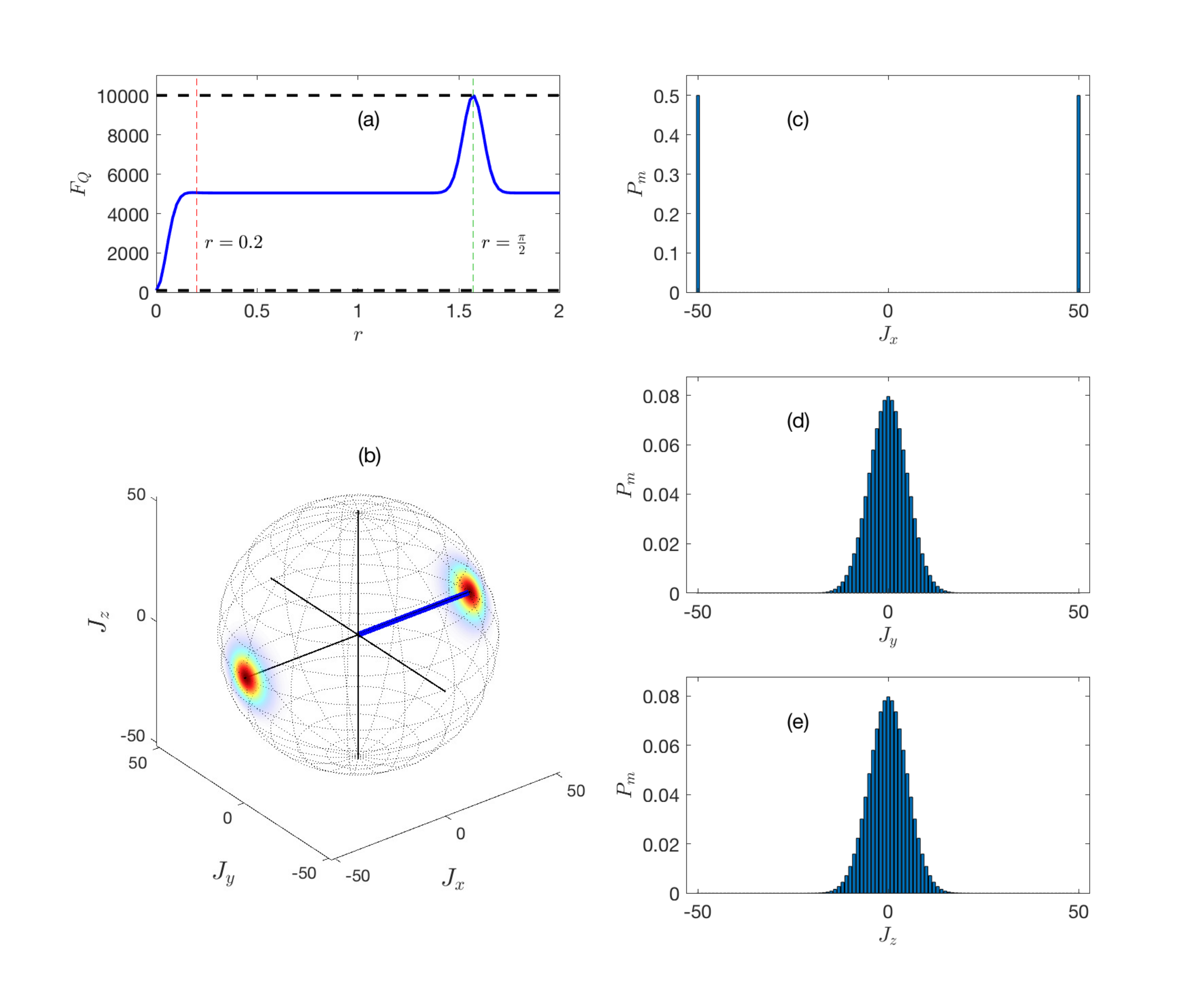}
		\caption{Properties of the cat state: $|\psi_1\rangle = \exp\left(i r \hat{J}_z^2\right)\exp\left(i \tfrac{\pi}{2} \hat{J}_y\right) |\tfrac{N}{2}\rangle$, for $r=\pi/2$. (a): The QFI is calculated via $F_Q = 4 \left(\langle \psi_1| \Jhat_n^2|\psi_1\rangle - |\langle \psi_1| \Jhat_n|\psi_1\rangle |^2\right)$, where $\Jhat_n$ is the pseudo-spin operator along which the QFI is maximum.  The lower and upper dashed black lines indicate the shot-noise limit ($F_Q = N$) and Heisenberg limit ($F_Q = N^2$), respectively.  (b) $Q(\theta, \phi)$. The direction of $\hat{J}_n$ is indicated by the thick blue line. (c-e): The probability distribution $P_m = |\langle m_j|\psi_1\rangle|^2$, where $m_j$ is the $m$th eigenstate of $\Jhat_j$, for $j = \{x,y,z\}$. }
		\label{catfig}
	\end{center}
\end{figure}

\subsection{QPT}
The QPT state was generated via evolution by a time-dependent Hamiltonian of the form 
\begin{equation}
\hat{H} = \hbar \chi (\Jhat_x \cos^2 \tfrac{\pi}{2} \tfrac{t}{t_0} + \Jhat_z^2 \sin^2 \tfrac{\pi}{2} \tfrac{t}{t_0}), 
\end{equation}
such that 
\begin{equation}
U_1 = \left(\mathcal{T} \exp \left( \tfrac{-i}{\hbar}\int_0^{t_0} \hat{H}(t^\prime) dt^\prime \right)\right) \exp(i \frac{\pi}{2}\Jhat_y) , 
\end{equation}
where $\mathcal{T}$ represents the time-ordering operator. In the limit $\chi t_0\rightarrow \infty$, $U_1|\tfrac{N}{2}\rangle = |0\rangle$, the twin-Fock state. We chose a moderate value $\chi t_0 = 20$, such that the final state contains non-zero elements on either side of  $m=0$. The Husimi $Q$-function, probability distribution, and QFI for this state are shown in Fig.~(\ref{QPTfig}).

\begin{figure}[h!!!]
	\begin{center}
		\includegraphics[width=\textwidth]{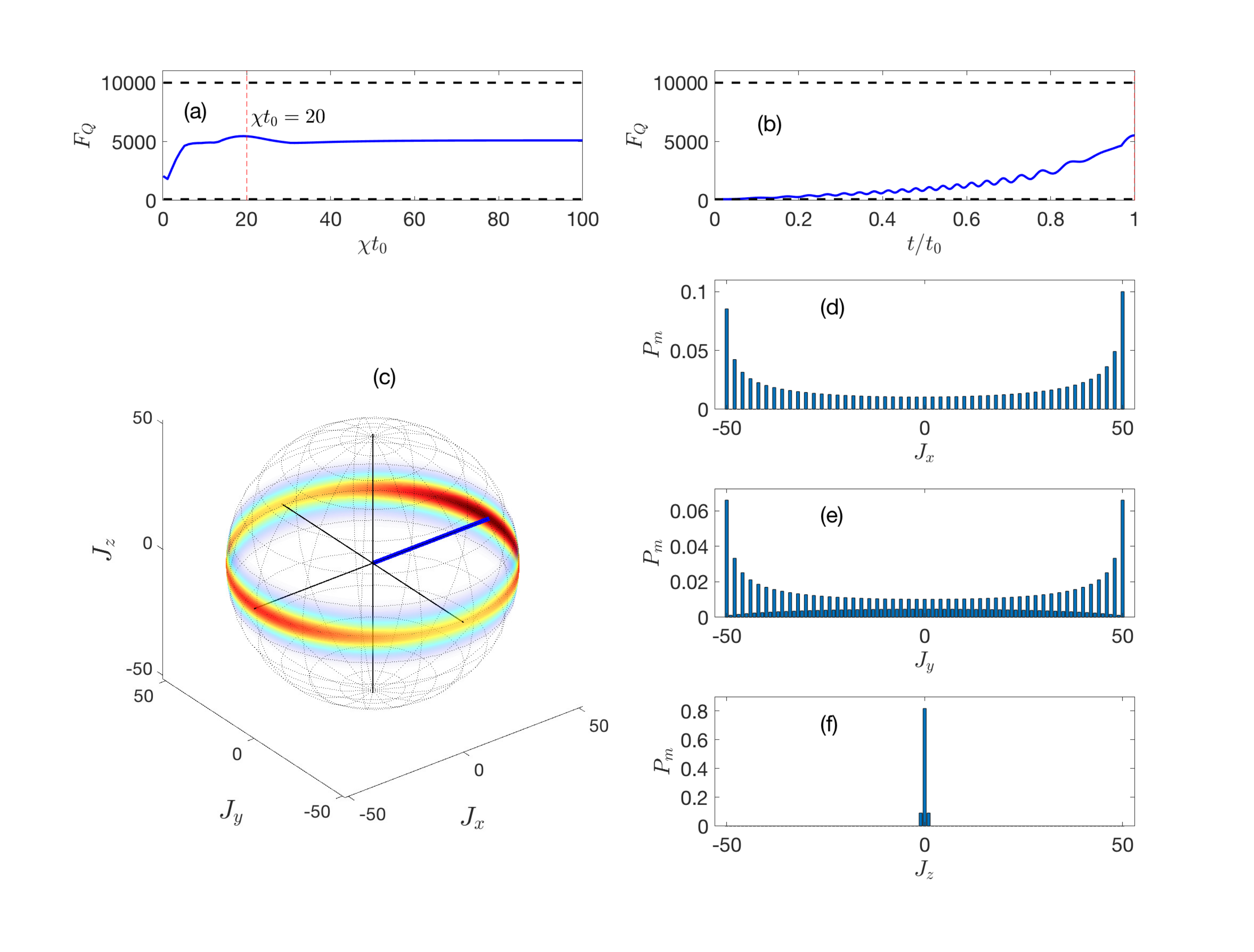}
		\caption{Properties of the QPT state: $|\psi_1\rangle = U_1 = (\mathcal{T} \exp ( \tfrac{-i}{\hbar}\int_0^{t_0} \hat{H}(t^\prime) dt^\prime ) ) \exp(i \frac{\pi}{2}\Jhat_y) |\tfrac{N}{2}\rangle$, where $\hat{H} = \hbar \chi (\Jhat_x \cos^2 \tfrac{\pi}{2} \tfrac{t}{t_0} + \Jhat_z^2 \sin^2 \tfrac{\pi}{2} \tfrac{t}{t_0})$.  (a): The QFI is calculated via $F_Q = 4 \left(\langle \psi_1| \Jhat_n^2|\psi_1\rangle - |\langle \psi_1| \Jhat_n|\psi_1\rangle |^2\right)$, where $\Jhat_n$ is the pseudo-spin operator along which the QFI is maximum, as a function of $\chi t_0$, where $t_0$ is the maximum time. (b): The QFI as a function of $t/t_0$, for $\chi t_0 = 20$. In (a) and (b), the lower and upper dashed black lines indicate the shot-noise limit ($F_Q = N$) and Heisenberg limit ($F_Q = N^2$), respectively.  (c) $Q(\theta, \phi)$. The direction of $\hat{J}_n$ is indicated by the thick blue line. (d-f): The probability distribution $P_m = |\langle m_j|\psi_1\rangle|^2$, where $m_j$ is the $m$th eigenstate of $\Jhat_j$, for $j = \{x,y,z\}$. }		
		\label{QPTfig}
	\end{center}
\end{figure}

\subsection{QND}
The QND state is was selected as a mixture of $\Jhat_z$ eigenstates. Specifically
\begin{equation}
\rho = \frac{\sum_m \exp\left({-\frac{m^2}{\Delta^2}}\right)|m\rangle\langle m| }{\left(\sum_m e^{-\tfrac{m^2}{\Delta^2}}\right)} \, .
\end{equation}
In order to calculate the QFI of a mixed state, we must use $F_Q = \mathrm{Tr}[\rho (\hat{L}[\rho])^2]$, where $\hat{L}$ is the symmetric logarithmic derivative. For our case, the QFI takes the from
\begin{equation}
F_Q = \sum_{i,j} \frac{2|\langle e_i|\hat{J}_n|e_j\rangle|^2(\lambda_i-\lambda_j)^2}{\lambda_i + \lambda_j} \, \label{simlogdiv}
\end{equation}
where $\rho|e_j\rangle = \lambda_j |e_j\rangle$. For our state, $\Jhat_n$ lies in the $x-y$ plane, so for definitiveness we chose $\Jhat_n = \Jhat_y$. The Husimi $Q$-function, probability distribution, and QFI for this state are shown in Fig.~\ref{QNDfig}. 

\begin{figure}[h!!!]
	\begin{center}
		\includegraphics[width=\textwidth]{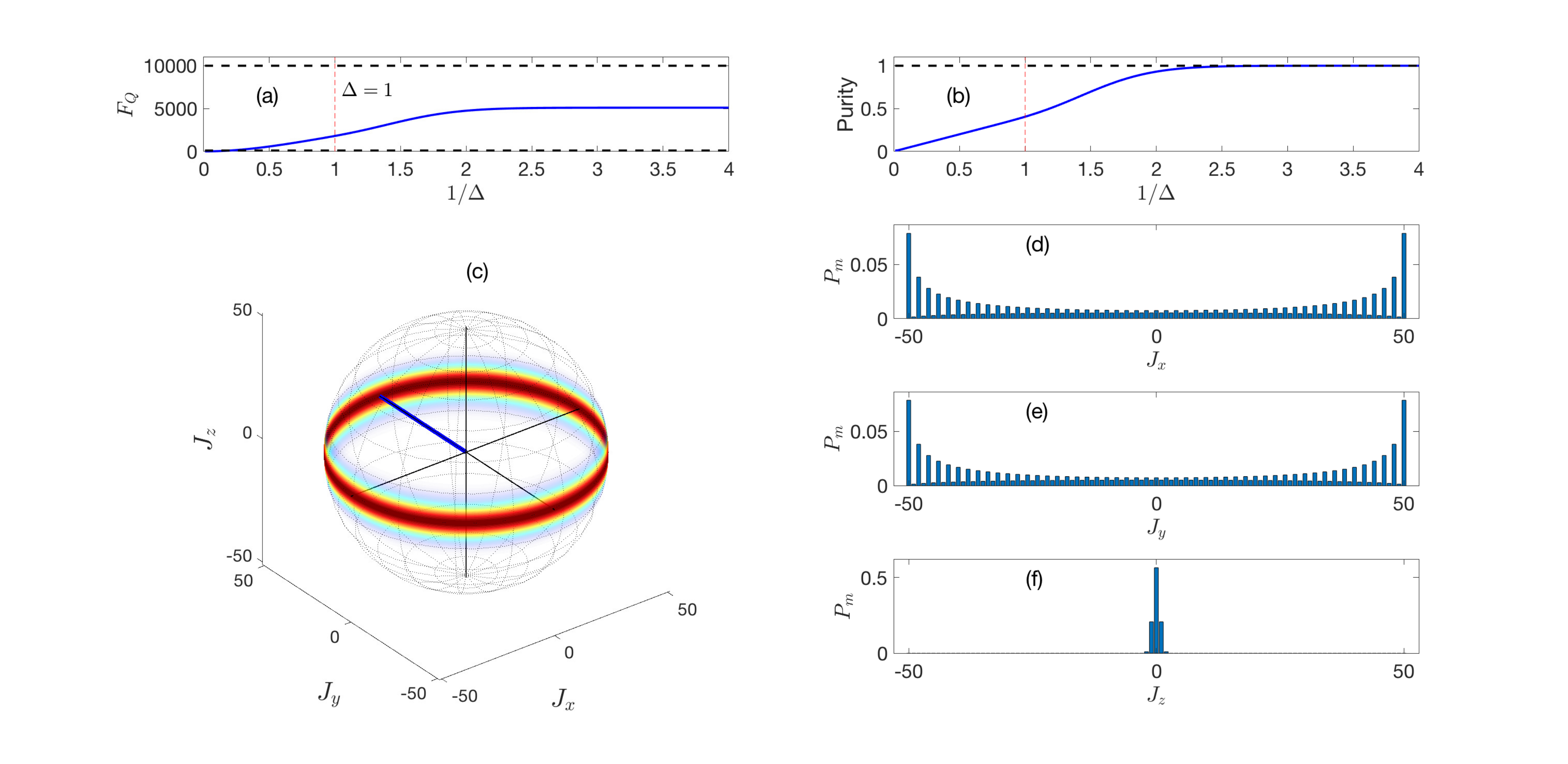}
		\caption{Properties of the QND state: $\rho = \sum_m e^{-\frac{m^2}{\Delta^2}}|m\rangle\langle m| /(\sum_m e^{-\tfrac{m^2}{\Delta^2}})$, for $\Delta=1$. (a): The QFI is calculated via \eq{simlogdiv}.  The lower and upper dashed black lines indicate the shot-noise limit ($F_Q = N$) and Heisenberg limit ($F_Q = N^2$), respectively. (b): The purity $\gamma = \mathrm{Tr}[\rho^2]$.  (c) $Q(\theta, \phi)$. For our state, $\Jhat_n$ lies in the $x-y$ plane, so for definitiveness we chose $\Jhat_n = \Jhat_y$ (indicated by the thick blue line). (d-f): The probability distribution $P_m = |\langle m_j|\psi_1\rangle|^2$, where $m_j$ is the $m$th eigenstate of $\Jhat_j$, for $j = \{x,y,z\}$.}
		\label{QNDfig}
	\end{center}
\end{figure}

\subsection{Coherent Spin State}
For completeness, we consider the coherent spin state given by $|\psi_1\rangle = U_1 |\tfrac{N}{2}\rangle$, where 
\begin{equation}
U_1 = \exp\left(i \tfrac{\pi}{2} \hat{J}_y\right). 
\end{equation}
Fig.~(\ref{CSSfig}) shows $F_C(\sigma)$ for the different IBRO. We see the same general trend as throughout the rest of the paper, except that $U_2 = U_\theta$ and $U_2 = U_1$ are identical.

\begin{figure}[h!!!]
	\begin{center}
		\includegraphics[width=\textwidth]{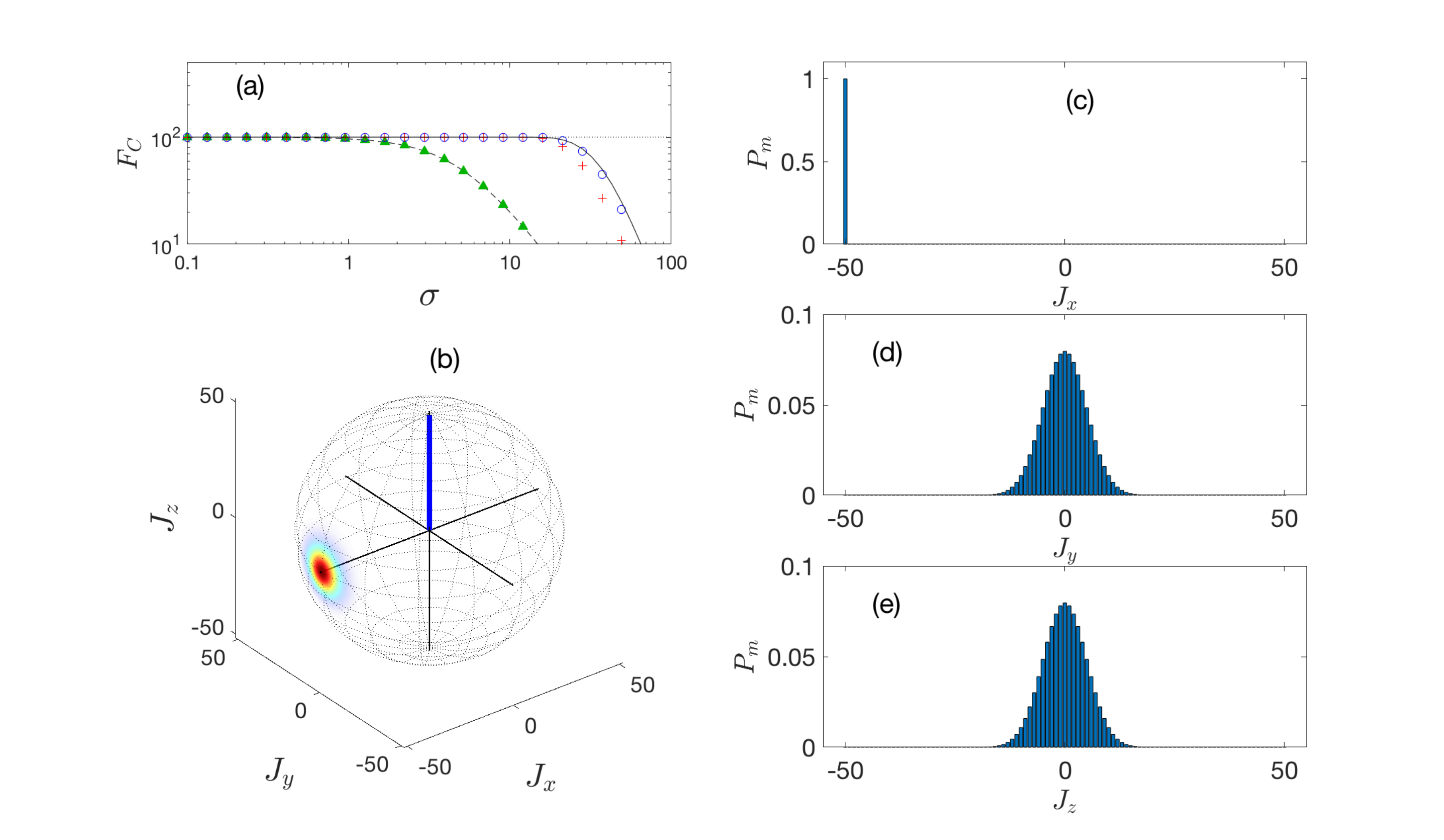}
		\caption{Properties of the CSS state: $|\psi_1\rangle = \exp\left(i \tfrac{\pi}{2} \hat{J}_y\right)|\tfrac{N}{2}\rangle$. (a): $F_C(\sigma)$ for $U_2 = U_\theta$ (dashed line), $U_2 = U_1$ (green triangles), $U_2 = U_\mathrm{flip}U_1$, (red plus symbols), $U_2 = U_\mathrm{opt}$ (blue circles), and the NQCRB (black solid line).  The QCRB is identical to the SNL ($F_Q = N$), indicated by the dotted line.  (b) $Q(\theta, \phi)$. For our state, $\Jhat_n$ lies in the $x-z$ plane, so for definitiveness we chose $\Jhat_n = \Jhat_z$ (indicated by the thick blue line). (c-e): The probability distribution $P_m = |\langle m_j|\psi_1\rangle|^2$, where $m_j$ is the $m$th eigenstate of $\Jhat_j$, for $j = \{x,y,z\}$.}
		\label{CSSfig}
	\end{center}
\end{figure}

\end{widetext}

\end{document}